\documentclass[
reprint,
]{revtex4-1}

\usepackage{graphicx}
\usepackage{dcolumn}
\usepackage{bm}

\begin{document}


\title{How to reconcile Information theory and Gibbs-Herz entropy
for inverted populated systems}


\author{Alessio Gagliardi$^{1}$, Alessandro Pecchia$^{2}$}
\affiliation{(1)  Technische Universit\"{a}t M\"{u}nchen,
Electrical Eng. and Information Tech., Arcisstr. 21 80333 M\"{u}nchen (Germany). alessio.gagliardi@tum.de\\
Tel: (+49) 89 289 26931.\\
}%
\affiliation{(2) CNR, Via Salaria Km 29,600, 0017 Monte Rotondo (Italy) \\
}%

\date{\today}

\begin{abstract}
In this paper we discuss about the validity of the Shannon entropy
functional in connection with the correct Gibbs-Hertz probability
distribution function. We show that there is no contradiction in
using the Shannon-Gibbs functional and restate the validity of
information theory applied to equilibrium statistical mechanics.
We show that under these assumptions, entropy is always a monotone
function of energy, irrespective to the shape of the density of
states, leading always to positive temperatures even in the case
of inverted population systems. In the second part we assume the
validity of the Shannon entropy and thermodynamic temperature,
 $T=dE/dS$, extended to systems under non-equilibrium steady state.
Contrary to equilibrium, we discuss the possibility and meaning of
a negative temperature in this case. Finally we discuss on Carnot
cycles operating with a non-equilibrium bath possessing a negative
temperature and leading to apparent efficiencies larger than one,
due to a wrong accounting af all the energy and entropy fluxes
present in the system, including the external driving forces.

\end{abstract}

\maketitle


\section{Introduction}

Recently a large debate has started about the existence of systems
- especially systems with upper bounded density of states that can
lead to inverted population conditions even in equilibrium - that
turns to have negative temperatures \cite{science}. The idea of a
negative temperature is an old concept, elaborated by many authors
\cite{negtemp1,negtemp2,negtemp3}. It comes from the possibility
that the derivative of entropy, $S$, with respect to internal
energy, $E$, exhibits a negative slope hence, from its
thermodynamical definition, temperature becomes negative.

In their paper \cite{nature} Dunkel et al. showed that this is not
correct and that the entire idea of negative temperature steams
from the wrong assumption about the entropy functional
(Boltzmann-Gibbs-Shannon) for a system in a microcanonical
ensemble. Subsequently, another paper \cite{hanngi} has been
published to strengthen this point, fundamentally concluding with
these three important findings:
\begin{itemize}
    \item A system which stays in an inverted population as
    equilibrium state must be in a microcanonical ensemble;
    \item The rigorous entropy functional is the Gibbs-Hertz
    entropy functional;
    \item For this choice of the entropy the temperature computed
    as derivative of entropy w.r.t. internal energy remains always
    positive.
\end{itemize}

In particular, their conclusion is also that the entropy
functional has not the shape of an information entropy functional
represented by the Shannon entropy and thus information
theoretical concepts should be carefully considered before being
applied to thermodynamics. They even suggest that the Gibbs-Hertz
entropy should be included as a "new" information measure if we
want to include the microcanonical ensemble within the formalism
of information theory.

Although their derivation is completely right, we demonstrate in
this work that it does not exclude an information theoretical
approach to statistical physics. Furthermore we show that the
Shannon standard entropy functional works and represents a general
functional, at least valid in equilibrium condition and in
non-equilibrium steady state (NESS), hence there is no need to
invent a new information measure. Finally, in the last part, we
briefly discuss under which conditions a negative temperature is
still possible and the implications on Carnot cycle efficiencies.
However, the last part does not include rigorous general
demonstrations, but rather a collection of several results and
then should be regarded as a suggestion for future investigations.

\section{Entropy: which functional?}

Before starting we make the same assumptions as in \cite{hanngi},
namely that the system is strictly isolated (microcanonical
ensemble), and is described by an Hamiltonian, $H(\xi,Z)$, where
$\xi$ denotes the microscopic state and $Z$ = ($Z_{1}$, ...)
comprises external control parameters. The system conserves energy
$E$ and the energy is bounded from below $E \geq 0$. The
probability distribution function (PDF) over the density of states
(DOS) of the phase space is assumed to be uniform over a subset of
the possible states $\xi$ of the system. However, this leaves an
uncertainty about the domain on which this uniform distribution is
different from zero. In the literature several possible choices
are discussed, the microcanonical density operator (MDO) defined
as:
\begin{equation}\label{rho1}
    \rho_{M}(\xi|\bar{E}, Z) = \frac{\delta(\bar{E} - H(\xi|Z))}{\omega(\bar{E},Z)},
\end{equation}
or the following:
\begin{equation}\label{rho2}
    \rho_{C}(\xi|Z; \bar{E}) = \frac{\theta(\bar{E} - H(\xi|Z))}{\Omega(\bar{E},Z)},
\end{equation}
that will be named cumulant density operator (CDO) where $\bar{E}$
is a parameter. In the former $\omega$ is defined as:
\begin{equation}\label{rho3}
    \omega(\bar{E},Z) = Tr[\delta(\bar{E} - H(\xi,Z))],
\end{equation}
where in the latter $\Omega$ is:
\begin{equation}\label{rho4}
    \Omega(\bar{E},Z) = Tr[\theta(\bar{E} - H(\xi,Z) )],
\end{equation}
with $\theta$ the Heaviside step function and $Tr$ is intended as
a trace of the corresponding quantum operator, or a normalizing
summation over all energy states.

If we use these two different density operators to get the PDF
over the DOS of the phase space, fixed the internal energy control
parameter, $\bar{E}$, we get two uniform distributions:
\begin{equation}\label{rho5}
   \rho_{M}(\xi|\bar{E},Z) \rightarrow \frac{1}{|\Gamma_{E = \bar{E}}|}
   \equiv \frac{1}{|\Gamma_{B}|},
\end{equation}
\begin{equation}\label{rho6}
   \rho_{C}(\xi|Z; \bar{E}) \rightarrow \frac{1}{|\Gamma_{E \leq \bar{E}}|}
   \equiv \frac{1}{|\Gamma_{GH}|},
\end{equation}
where the subscripts $B$ and $GH$ stand for Boltzmann and
Gibbs-Hertz. The difference is that the Boltzmann PDF is uniform
over all the states of the system with energy equal to $\bar{E}$,
while the Gibbs-Hertz is uniform over all states with energy $E
\leq \bar{E}$. If we insert the Boltzmann PDF inside the standard
Shannon entropy functional,
\begin{equation}\label{rho7}
    S = -k_{B} Tr[ \rho \ln \rho],
\end{equation}
where k$_{B}$ is the Boltzmann constant and $\rho$ the density
operator we get the Boltzmann entropy,
\begin{equation}\label{rho8}
    S_{B} = \ln |\Gamma_{B}| = \ln \epsilon\omega(\bar{E}).
\end{equation}
As pointed out in \cite{hanngi}, an infinitesimal $\epsilon$ is
actually needed in order to give a "volume" to the shell of states
with energy just equal to $\bar{E}$. It is important to stress
that in equilibrium and without any dissipation, under ergodic
assumptions, these are the states effectively explored by the
system during time evolution.

If the Gibbs-Hertz PDF is inserted in the Shannon functional,
 we get the correct Gibbs-Hertz entropy, as presented in \cite{hanngi}:
\begin{equation}\label{rho9}
    S_{GH} = \ln |\Gamma_{GH}| = \ln \Omega(\bar{E}).
\end{equation}
Thus we observe that the correct Gibbs-Hertz entropy for the
microcanonical ensemble does not need the invention of a novel
entropy functional, beyond the conventional Shannon entropy. What
is really needed is the correct PDF.

Information Theory \cite{cover} provides a perfectly consistent
framework inside which thermodynamics can be developed, but it
lacks an important piece of knowledge that can only come from
physical considerations. Information Theory is based on
functionals, such as the Shannon entropy \cite{note1}, from which
the entire theory is derived. For example two of the main
achievements of information theory, namely channel capacity and
rate distortion theory, are nothing else than respectively
maximizing or minimizing a mutual information functional. However,
information theory has nothing to say about the shape of the PDF,
the basic ingredient of the entropy functional. The theory just
predicts some statistical/probabilistic aspects of the system once
its PDF is known or assumed. The only constrain is that the PDF
has to be properly normalized, as any PDF should be, but this
leaves infinite classes of functions fulfilling this requirement.

Thus, information theory requires a correct PDF that can only be
obtained by empirical considerations, experimental evidences or
numerical simulations based on the underlying microscopic physics.

In the case of a system in a microcanonical ensemble, physical
considerations based on the fulfillment of the laws of
thermodynamics and the equipartition theorem, bring to conclude
that the correct PDF is the Gibbs-Hertz and not the commonly used
Boltzmann PDF, although the subtle difference between the two is
not usually perceived since in normal systems, for which the
density of states grows monotonically with energy, all the
contribution to the Gibbs-Hertz entropy is given by states with
energy $E=\bar{E}$, essentially recovering the Boltzmann result
\cite{kubo}. This becomes asymptotically correct in the
thermodynamic limit where the number of states with energy $E <
\bar{E}$, for a monotonic DOS, becomes negligible, so that
$\Omega$ converges to $\omega$ and the Gibbs-Hertz entropy becomes
the Boltzmann entropy. Several alternative forms of PDF are
discussed in \cite{hanngi}, where it is shown that all fall into
some contradiction when compared to physical reality. However,
this asymptotic equivalence is only an approximation which must be
taken with caution and forbids general conclusions. Among the
others, systems where Boltzmann MDO fails are small systems, for
which the thermodynamic limit does not hold \cite{note2} or those
characterized by an anomalous DOS, i.e., DOS that are not simply
growing monotonically with energy, but for instance have a maximum
and then decrease to zero.

However, this rises an important question: why the CDO is correct,
considering that a closed system without dissipation can only
"explore" states with $E = \bar{E}$?

A practical answer could be that the CDO is the one which gives
the right PDF and the correct (Gibbs-Hertz) thermodynamic entropy
and any further consideration could be put aside. The
thermodynamic laws should dictate the correct PDF and not any
other assumption.

A deeper understanding necessarily involves the fact that a
perfectly conservative system is only an idealization of reality.
Equilibration itself requires dissipation. The MDO (leading to
Boltzmann PDF), by sampling only states with $E = \bar{E}$, lack
any information about the thermodynamic inequivalence between
higher and lower energy states involved in the relaxations. As
pointed out correctly in \cite{hanngi} "A system occupying the
ground state can be easily heated (e.g., by injecting a photon),
whereas it is impossible to add a photon to a system in the
highest energy state. The Gibbs-Hertz entropy reflects this
asymmetry in a natural manner".

For the moment let's think about a simple anomalous DOS starting
at $E=0$ with DOS(0) = 0, reaching a maximum and then going to
zero, and let assume that our system is in a microcanonical
ensemble. It is clear that if we use the Boltzmann PDF and
substitute it into the Shannon entropy we get a non monotonous
entropy. Computing the system temperature using the traditional
expression,
\begin{equation}\label{temp1}
    \frac{1}{T} = \frac{\partial S}{\partial E},
\end{equation}
leads to an apparently negative temperature, whenever $\bar{E}$ is
larger than the DOS maximum. This has started a debate in the
literature about the interpretation of this result trying to give
significance to negative temperatures in general. However this
result is wrong.

The problem is precisely that the Boltzmann PDF assumes a sort of
symmetry between low and high energy states as it depends only on
the DOS value. In practice there is no difference if a system is
in a high or low energy with respect to the DOS. Instead, by using
the correct Gibbs-Hertz PDF one gets a monotonous entropy,
regardless the shape of the DOS. A direct consequence is that the
temperature remains always positive.

There is a last point to clarify. If the correct microcanonical
PDF is that of Gibbs-Hertz, why in many averaged quantities we use
safely the Boltzmann PDF? The answer is simply given by analyzing
how the averages are taken (see for example eq. 11 in
\cite{hanngi} associated to the equipartition theorem):
\begin{equation}\label{TGH1}
    T_{GH} = \left \langle \xi_i \frac{\partial H}{\partial \xi_i}  \right
    \rangle_{E = \bar{E}},
\end{equation}
where $\xi_i$ is any of the $i^{th}$ degrees of freedom describing
the state $\xi$, the subscript $GH$ stresses the fact that
temperature is computed using the derivative of the Gibbs-Hertz
entropy and the brackets $\langle ... \rangle$ means ensamble
average. We observe that the average is taken under the constrain
$E = \bar{E}$ and we notice that this is equivalent to saying that
the average is taken over a conditioned PDF of the original
Gibbs-Hertz PDF. In fact, it is trivial to see that we obtain the
MDO from the CDO (and the Boltzmann PDF from the Gibbs-Hertz PDF)
just by making the conditioning,
\begin{equation}\label{xi}
    \rho_{M}(\xi) = \rho_{C}(\xi|E = \bar{E}).
\end{equation}
In practice we can drop the redundant symbol, $E = \bar{E}$, at
the averaging brackets $\langle ...\rangle$, as it is redundant
when using the conditioned MDO in the average. This subtle result,
coupled to the asymptotic equivalence between Boltzmann and
Gibbs-Hertz PDFs in the thermodynamic limit, has been the source
of a large confusion and a widespread use of the Boltzmann PDF
taken as generally valid.

The major conclusion of this section is that the Shannon entropy
functional for the thermodynamic entropy seems to be valid in more
general cases, provided the correct PDF is used. We have also seen
that the Boltzmann PDF is an asymptotic approximation of the
Gibbs-Hertz PDF. Furthermore the Boltzmann PDF is correct when
used in averaged quantities because it is nothing else that the
conditioned Gibbs-Hertz PDF on the surface $E = \bar{E}$. However,
as we will show in the next part of this work, it is possible to
have genuine situations where the temperature becomes negative,
but this requires non-equilibrium conditions.


\section{Negative temperature, dark energy and Carnot cycle efficiency}

This section is more speculative, but we believe it can help to
clarify several aspects concerning inverted populated systems and,
more generally, all those indicating the existence of a negative
temperature and using the equipartition relation a consequent
"dark energy".

First of all we will enlist our position about the three different
topics and then explain in more detail:
\begin{itemize}
    \item Existence of negative temperature? YES;
    \item Dark Energy? NO;
    \item Carnot cycles with efficiency larger than unity? TAKEN VERY CAREFULLY.
\end{itemize}

In order to explain our position we will mention the work made in
\cite{indiani}. In their paper the authors discussed the
thermodynamics of a particle in a noisy environment described by a
Langevin equation. They assumes that the particle starts with an
energy larger than the equilibration energy and analyze the
exchange of heat and work between the particle and the
environment.

It is clear that such a system describes a particle in a canonical
ensemble. They show that until the particle is out of equilibrium,
its state can be described by two different temperatures, one
simply related to the particle kinetic energy, $k_B T_{ET}\propto
\bar{E}$, and the other computed using $T_{ED}=dE/dS$. The first
temperature is fundamentally linked to the average energy per
degree of freedom of the system and thus is always positive and
related to the equipartition theorem. This temperature definition
is always possible even under non-equilibrium conditions, since
the average energy per degree of freedom is always a well-defined
quantity. The second temperature is a thermodynamic temperature,
that requires stretching the concept of entropy also to systems
out of equilibrium and in many cases this extension is still
questionable. In their work Narayanan and Srinivasa assume that
the form of the Shannon is still valid in the case of a non
equilibrium steady state (NESS), in which a distribution function
is meaningful. Then, they link $dE/dS$ with the average curvature
of the PDF, $kT_{ED}=\langle \nabla H \nabla H \rangle / \langle
\nabla ^2 H \rangle$, where $H$ is the Hamiltonian of the system,
leading, via the De Brujin's identity, to an expression related to
the Fisher information \cite{indiani}.

We do not want to discuss here the general validity of this
approach, but we find it elegant and meaningful. Assuming this is
correct, at least for the particular system analyzed there, we
want to concentrate on its consequences.

Calling $T_{0}$ the environment temperature, it is important to
notice that in equilibrium $T_{ED} = T_0 = T_{ET}$. This result,
apparently obvious, is in fact quite special to the higher
constrain of the equilibrium state \cite{touchette,derrida}, and
explains why temperature is such a useful concept in equilibrium.
It also explains why temperature is still useful in the
microcanonical ensemble, where it is not a control parameter.

To this result we add the following finding. Consider the average
energy, $\langle E \rangle = \sum_i p_i E_i$, and entropy, $S =
-k_B \sum_i p_i \ln p_i$, and a small perturbation to the
distribution probability of the general form
\begin{equation}\label{peps}
 p'_i = p_i + \epsilon \sum_k \pi_{ik} p_k,
\end{equation}
where $\epsilon$ is infinitesimal and $\pi_{ik}$ a general
convolution matrix, subject to $\sum_i \pi_{ik} p_k = 0$,
necessary to preserve the condition $\sum_i p'_i = 1$. Assuming
further that the perturbation on $S$ and $\langle E \rangle$ does
not perturb the state energies, $E_i$, and that the equilibrium
distribution is the canonical $p_i = exp(-\beta E_i)/Z$, it is not
difficult to find that
\begin{equation}
 \frac{dS}{d \langle E \rangle}= \frac{\delta S}{\epsilon} \frac{\epsilon}{\delta \langle E \rangle} = \frac{1}{T},
\end{equation}
completely independent on the form of the perturbation $\pi_{ik}$.
This shows that in equilibrium conditions, no matter how physical
thermometers couple to a system, they will all agree in measuring
the same temperature.

Not surprisingly, under non-equilibrium conditions, we have in
general $T_{ED} \ne T_{ET} \ne T_0$. What does this mean? It
simply means that in non-equilibrium conditions the entire concept
of temperature must be taken with care. Different ways to compute
temperature are connected to different aspects of the
thermodynamics of the system under consideration and a general
definition is probably not possible. While $T_{ET}$ is linked to
the energy content of the system, $T_{ED}$ is related to a
curvature of the PDF in phase space. An interesting result of
\cite{indiani} is that the amount of work and heat exchanged
between the particle and the environment are related
(proportional) to temperature differences as,
\begin{equation}\label{DQ}
    \Delta Q \propto (T_{0} - T_{ET}),
\end{equation}
\begin{equation}\label{DW}
    \Delta W \propto (T_{ED} - T_{ET}).
\end{equation}
Despite this result lack generality, we think the reasoning gives
useful insights and these final expressions are noteworthy,
showing that different temperatures can be used in a thermodynamic
sense to define the direction of heat and energy fluxes.

Now let's focus on the problem of an inverted populated systems in
steady state. We can consider two possibilities: a) the system is
in an inverted populated state, but it is in equilibrium
conditions, b) the system is in a NESS forced by some external
drive \cite{udo}.

In the first case we must have $T_{ED}$ = $T_{ET} > 0$. This
condition is the one discussed in \cite{hanngi} by Hilbert and
coworkers and the conclusion is that the system must be in a
microcanonical ensemble and the temperatures is unique and
positive. A corollary is that no dark energy exists and the
efficiency of any Carnot cycle where one of the bath is an
inverted populated system is bound to $\eta \leq 1$, with:
\begin{equation}\label{eta}
    \eta = 1 - \frac{T_{cold}}{T_{hot}}.
\end{equation}
We do not further discuss this case as all the conclusions in
\cite{hanngi} are correct and widely explained there.

When there is an external drive or a feedback machine,  forcing
 the system out of equilibrium, then, in general,
\begin{itemize}
    \item $T_{ED} \ne T_{ET}$;
    \item The system is not in a microcanonical ensemble
    as an external feedback has to interact with the system in order to
    preserve its NESS.
\end{itemize}
In this case there are no fundamental reasons preventing $T_{ED}$
to become negative since the drive or the feedback can push the
system in a state where the PDF inserted into the Shannon entropy
leads to an entropy function non monotonous with energy.

As we have previously discussed, in the equilibrium this is
forbidden as there is an intrinsic thermodynamic difference
between occupying low or high energy states. We can think that an
external drive, making work, can circumvent this asymmetry leading
to a negative $T_{ED}$. However, we notice that if $T_{ED} < 0$,
since $T_{ET} > 0$ (always), necessarily $T_{ED} \neq T_{ET}$. We
also notice that using the results of \cite{hanngi}, the special
condition $T_{ED} < 0$ is only possible under non-equilibrium
conditions. The final observation of this discussion is that
although $T_{ED} < 0$ may have a meaning, this temperature is
related to entropy and state distribution but is generally
unrelated to the energy content of the system, ruling out the
existence of dark energies.

Let's now discuss the Carnot cycle efficiency of a cycle between
two reservoirs in equilibrium, one of which has an inverted
population condition. We call $R_{st}$ the standard reservoir,
$R_{inv}$ the reservoir with inverted population and $S_{1}$ the
system coupling in alternation with the two reservoirs along the
Carnot cycle.

Also in this case, for the inverted population reservoir, we
consider the two possibilities: equilibrium/microcanonical or
NESS. If $R_{inv}$ is in thermodynamic equilibrium, then (assuming
as usual a weak coupling to $S_1$), it must be characterized by
$T_{inv}>0$, hence $\eta = 1 - T_{inv}/T_{st} \leq 1$. In the case
of a NESS and $T_{ED}<0$, then $R_{inv}$ must be under
non-equilibrium conditions. Apparently in this case we may
conclude $\eta>1$, however the traditional definition of Carnot
efficiency, equation (\ref{eta}), must be taken with care in this
case. Even assuming (\ref{eta}) is still correct, we have already
concluded that under these conditions $T_{ED} \ne T_{ET}$,
therefore it is not clear which temperature must be used in
(\ref{eta}). Although there are indications that the correct
Carnot efficiency can be obtained for a NESS system using $T_{ED}$
\cite{alessio}, this result is not general, and it has been shown
only for the case $T_{ED}>0$ and may not apply to all NESS
systems, especially those with a negative $T_{ED}$. In fact the
two reservoirs $R_{st}$ and $R_{inv}$ and the system $S_1$ cannot
make an isolated engine since an external drive or a feedback is
certainly needed to keep $R_{inv}$ in its inverted NESS. Any such
driver is making work on the system or producing a large amount of
entropy to accomplish this task. In accounting the total Carnot
efficiency, we cannot forget this work. Therefore, the surprising
result $\eta>1$ is meaningless. This simply means that in doing
the energy accounting we are disregarding the energy flux between
driver and $R_{inv}$, which continuously makes work on $R_{inv}$,
increasing its own entropy to keep $R_{inv}$ into the inverted
state. The entropy production associated makes the total
efficiency to inevitably drop below unity. So, unfortunately, no
dark energy and perpetual motion are possible.

\section{Conclusion}

In this paper we have tried to clarify some points about the
debate between Boltzmann, Gibbs-Hertz and Shannon entropy and the
concepts of negative temperature, dark energy and Carnot
efficiencies larger than one in inverted populated systems.

We agree that in a system in equilibrium and in a microcanonical
ensemble the Gibbs-Hertz entropy is the correct thermodynamic
entropy, however this does not contradict the fact that Shannon
entropy is still the right functional for thermodynamic entropy in
general. It is just telling us that we have to choose the right
probability density function. In particular in the microcanonical
ensemble the correct one is the Gibbs-Hertz uniform density
function derived from the cumulant density operator. This is also
the right density function because it produces the entropy with
the correct "knowledge" of the directionality in energy state for
which high energy states are not equivalent to lower energy
states. Moreover, the Gibbs-Hertz entropy satisfies all the
thermodynamic laws and the equipartition theorem.

The source of the confusion is that for other constrained averages
in the microcanonical ensemble another probability density
function is used, the Boltzmann density, derived from the
microcanonical density operator, which incidentally becomes almost
equivalent to the Gibbs-Hertz density function for standard
density of states, especially in the thermodynamic limit. However,
the discussion of anomalous density of states \cite{hanngi} has
shown that the Gibbs-Hertz is correct and the Boltzmann is wrong
in more general cases. The Boltzmann uniform distribution is
anyway correct for all the averaging which require a conditioned
Gibbs-Hertz distribution, in particular when we require $E =
\bar{E}$, with $\bar{E}$ the energy of the system fixed as control
parameter. This shows that we should be careful in choosing the
correct density function for calculating different averages.

The second part of this work was focused on analyzing the
thermodynamics of inverted populated systems. We have pointed out
that inverted populated systems can exist in steady state only in
two possible conditions: equilibrium or non-equilibrium steady
state. In the first case, as demonstrated by \cite{nature}, the
system must be isolated (microcanonical ensemble), the correct
entropy functional is the Gibbs-Hertz which leads always to a
positive defined temperature. This forbids any dark energy,
negative temperature and it shows that if the system is used like
a reservoir in a Carnot engine leads to a maximal unitary
efficiency.

In the second case  (non-equilibrium steady state), we have shown
that there are evidences that point on the fact that temperature
is not a single value and different ways to compute this parameter
can lead to different values. In particular we can indeed keep the
system in a way to get a negative temperature obtained from the
derivative of entropy w.r.t. energy. However, this means that this
temperature has nothing to do with the temperature derived from
the equipartition theorem and in particular with the energy
content of the system. So again no dark energy can exist.

Concerning the Carnot efficiency in the presence of non
equilibrium steady state with an inverted population reservoir and
a negative temperature, we have argued that we should be careful
as the entire concept of Carnot efficiency equation cannot be
taken for granted. But even assuming it is correct, obtaining an
efficiency larger than one does not have any meaning, simply
because we are not taking into account the cost of keeping one of
the two reservoirs in non equilibrium conditions. Thus it would be
better to call such parameter with a different name, such as
"partial efficiency", in order to clearly distinguish it from the
real efficiency.



\begin{thebibliography}{99}
\bibitem{note2} There is in fact no strong reason why we must use
always thermodynamic limit as well as we must not necessarily
assume an infinite crystal in condensed matter physics and rely on
exact Bloch theorem.
\end{thebibliography}

\end{document}